\documentclass{PoS}

\newcommand{\preprintline}{\newline
\vskip -4.6cm
\rightline{\parbox{4cm}{\large\rm{ITEP-LAT/2005-19 \\ KANAZAWA/05-11}}}
\vspace{2.8cm}}

\PoS{PoS(LAT2005)307}

\title{\vskip 10mm
Towards SU(2) invariant formulation \\ of the monopole confinement mechanism\preprintline
}

\ShortTitle{Towards SU(2) invariant formulation of the monopole confinement mechanism}

\author{\speaker{Katsuya Ishiguro}\thanks{The authors thank RIKEN for their
        support of computer facilities and use of the RSCC computer clusters.}\\
        Institute for Theoretical Physics, Kanazawa University, Kanazawa
        920-1192, Japan\\
        and RIKEN, Radiation Laboratory, Wako 351-0158, Japan\\
        E-mail: \email{ishiguro@hep.s.kanazawa-u.ac.jp}}
\author{Yoshihiro Mori, Yoshifumi Nakamura, Toru Sekido and Tsuneo Suzuki\thanks{T.S. is supported
       by JSPS Grant-in-Aid for Scientific Research on Priority Areas 13135210 and (B) 15340073.}\\
        Institute for Theoretical Physics, Kanazawa University, Kanazawa
        920-1192, Japan \\
        and RIKEN, Radiation Laboratory, Wako 351-0158, Japan}
\author{M.~N.~Chernodub and M.~I.~Polikarpov\footnote{M.N.Ch. and M.I.P. are supported
        by  grants RFBR 04-02-16079, 05-02-16306, DFG 436 RUS 113/739/0 and MK-4019.2004.2.}\\
        Institute of Theoretical and Experimental Physics ITEP, 117259
        Moscow, Russia}
\author{V.~I.~Zakharov\\
        Max-Planck Institut f\"ur Physik, F\"ohringer Ring 6, 80805,
        M\"unchen, Germany}

\abstract{The type of the vacuum is studied numerically in the
maximally Abelian (MA) gauge and in the Landau (LA) gauge of
$SU(2)$ gluodynamics. The type of the vacuum is determined by a
ratio between the dual coherence and the dual penetration lengths.
The dual penetration length is determined from correlations
between Wilson loops and electric fields in both gauges. The dual
coherence length is found from correlations between Wilson loops
and dimension-2 operators both in the MA and the LA gauges. This
determination of the coherence length is supported by theoretical
and numerical observation that the dimension-2 gluon operators in
the studied gauges have a strong correlation with the monopole
current determined in the MA gauge. We find numerically that the
dual penetration lengths and the dual coherence lengths in the LA
and the MA gauges are almost the same. Therefore we conclude, that
in both gauges the type of the vacuum in the confinement phase is
near to the border between the type 1 and the type 2 dual
superconductors.}

\FullConference{XXIIIrd International Symposium on Lattice Field Theory\\
         25-30 July 2005\\
         Trinity College, Dublin, Ireland}

\newcommand{\beqn}{\begin{eqnarray}}
\newcommand{\eeqn}{\end{eqnarray}}
\newcommand{\eq}[1]{(\ref{#1})}

\begin{document}

\section{Introduction}
It is conjectured that the dual Meissner effect caused by the
monopole condensation is the color confinement
mechanism~\cite{ref:dual:superconductor}. The conjecture seems to
be realized if we perform Abelian projection in the maximally
Abelian (MA) gauge where Abelian component of the gluon field and
Abelian monopoles are found to be dominant~\cite{Reviews}.
Numerical calculations show that the vacuum of quenched $SU(2)$
QCD ($SU(2)$ gluodynamics) is near the border between the type 1
and the type 2 dual superconductor~\cite{ref:known:border},
although there are some claims that it is a superconductor of
weakly type 1, Refs.~\cite{bali98,Koma:2003hv}. Since the explicit
definition of a dual Higgs field is unknown the coherence length $\xi$
is usually calculated using classical Ginzburg-Landau equations,
while the penetration length $\lambda$ can be calculated directly measuring
the correlations between Wilson loops and (non-)Abelian
electric fields. Below we show that the coherence length $\xi$ can
be derived in the MA gauge also from the measurement of the
monopole density around a chromomagnetic flux.

The MA gauge is just one gauge among infinite possible
gauges. Since the physics should be gauge-independent, it is
important to know the confinement mechanism as well as the type of
the vacuum in another gauge. This problem has been discussed
recently in Ref.~\cite{suzuki05} where the Landau (LA) gauge is
considered and for Abelian components the dual Meissner effect is
observed. A magnetic displacement current plays the role of the
solenoidal supercurrent which squeezes the Abelian electric
fields. The observation of the dual Meissner effect in the LA
gauge suggests that there exists a gauge-independent definition of
the monopole condensation.

In order to fix the type of the vacuum in the MA and in the LA
gauges we use the following scheme. First, we demonstrate numerically
a strong correlation between the
dimension 2 gluon operators (namely, the operator
$A^+A^-(s)\equiv\sum_{\mu}[(A^1_{\mu}(s))^2+(A^2_{\mu}(s))^2]$ in
the MA gauge and $A^2(s)=A^+A^-(s)+A^3A^3(s)$ in the LA
gauge~\cite{fedor1}) and the monopole currents $|k_{\mu}(s)|$
(defined in the MA gauge). Then we show that the monopole
density is strongly correlated with the position of the QCD
string. Indeed, in the dual Ginzburg-Landau
model~\cite{suzuki88} of the SU(2) vacuum, the expectation value of
the (squared) monopole density around the string worldsheet $\Sigma$ is given by
the sum of the solenoidal current and the quantum correction, respectively~\cite{ref:basic:paper}:
\beqn
\langle k^2_\mu \rangle_\Sigma \equiv (k^{\mathrm{string}}_\mu)^2 +
(k^{\mathrm{quant}}_\mu )^2 = \left[\eta^2 m_B K_1(m_B \rho)\right]^2 +
\frac{g^2 |\Phi(\rho)|^4 \Lambda^2}{16 \pi^2} + \dots \nonumber\\
\to
\frac{g^2 \Lambda^2\, \eta^4}{16 \pi^2} \Biggl[1 - 4 \sqrt{\frac{\pi \xi}{2 \rho}} \, e^{ - \rho/\xi}\Biggr] + \dots
\quad [\mbox{in the limit } \ \rho \gg \xi]\,.
\label{eq:density:limit}
\eeqn
Here $m_B \equiv 1/\lambda$ is the mass of the dual gauge boson,
$\eta$ is the expectation value of the Higgs field, $\Lambda$ is
an UV-cutoff, and $\rho$ is the distance to the string worldsheet (the
string is taken to be infinitely long, straight and static for the
sake of simplicity). Moreover, the limit~\eq{eq:density:limit} demonstrates
that the leading behavior of the monopole density at large
distances is controlled by the coherence length $\xi$ and not by
the penetration length $\lambda$.

Thus, coherence lengthes in the MA and LA gauges can be found from
the correlation between the corresponding dimension 2 operators
and the Wilson loops. The comparison of the penetration length and
the coherence lengthes reveals that they are almost the same.
Consequently, we conclude that the vacuum is near the border
between the type 1 and the type 2 dual superconductors in the MA
gauge. Below we show the results of the numerical simulations which support our conclusion.

\newpage

\section{Numerical results}

\subsection{Method}
We use an improved gluonic action found by Iwasaki~\cite{iwasaki}
which was already implemented in Ref.~\cite{suzuki05}:
$
S = \beta \left\{C_0\! \sum {\mathrm{Tr}} \mbox{(plaquette)} + C_1\! \sum {\mathrm{Tr}} \mbox{(rectangular)} \right\}.
$
The mixing parameters are fixed as $C_0 + 8 C_1 = 1$ and $C_1=-0.331$.
We adopt the coupling constant $\beta=1.2$ which corresponds to
the lattice spacing $a(\beta=1.2) = 0.0792(2)$fm. The lattice size is $32^4$
and we use around  5000 thermalized configurations for measurements.
To get a good signal-to-noise ratio, the APE smearing technique~\cite{APE}
is used when evaluating Wilson loops
$W(R, T)=W^0+iW^a\sigma^a$.
The thermalized vacuum configurations are gauge-transformed in the MA($+$ U1LA)
gauge and in the LA gauge.

\subsection{MA gauge}

The MA gauge is defined by the maximization of the functional
$R[U] = \sum_l R_l[U]$, where $R_l[U] = \frac{1}{2}
{\mathrm{Tr}}[U_l \sigma_3 U_l^\dagger \sigma_3 ]$, with respect
to the SU(2) gauge transformations, $U^\Omega_{x,\mu} =
\Omega^\dagger_x U_{x,\mu} \Omega_{x+\hat \mu}$. In a naive
continuum limit one can make an identification of the dimension-2
operator $A^+_\mu A^-_\mu$ and a lattice quantity:
$A^+_\mu(x)A^-_\mu(x) = \frac{1}{2} (1 - R_{x,\mu}[U])$ where no
summation over $\mu$ is assumed.

\begin{figure}[htb]
\begin{minipage}{.48\textwidth}
\includegraphics[width=1.0\textwidth]{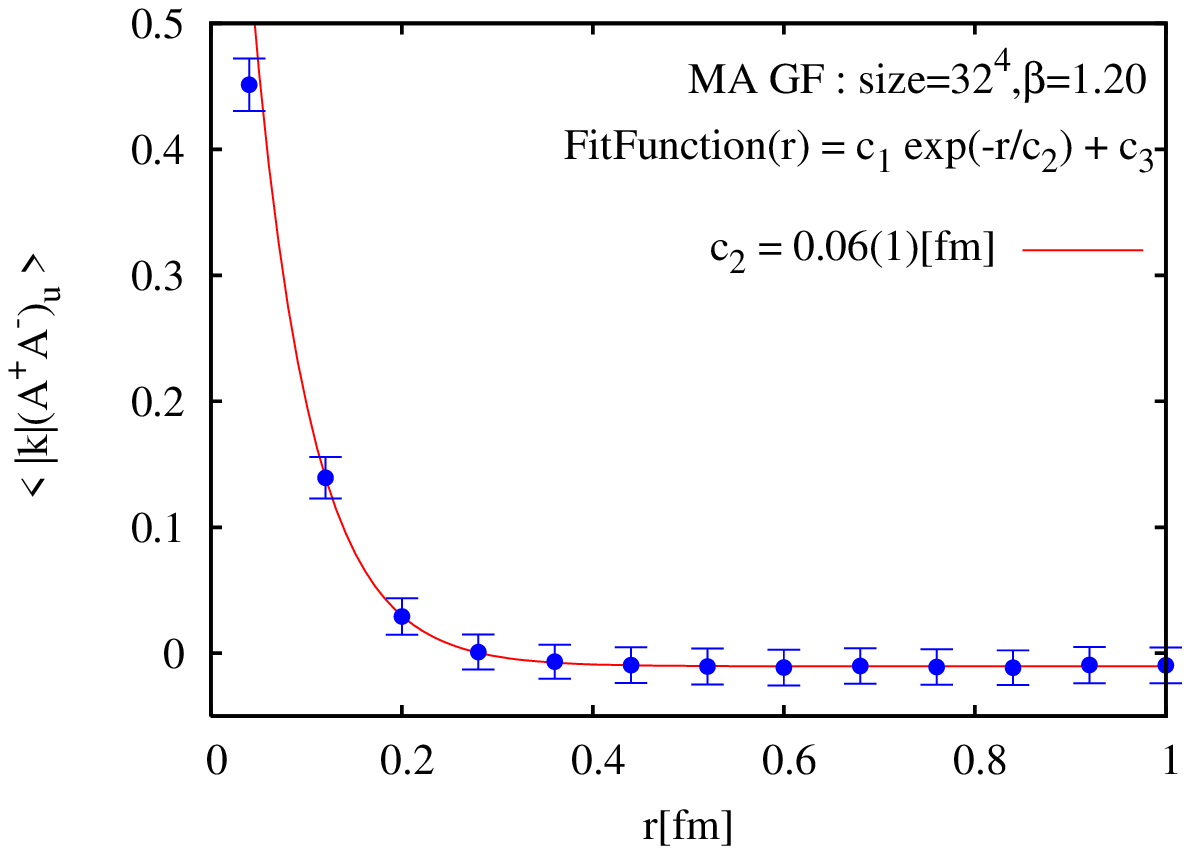}
\caption{\label{k-AA} The correlation between the monopole density
$|k_\mu|$ and the operator $A^+A^-$ in the MA gauge.
 The solid line denotes the best exponential fit.}
\end{minipage}
\hspace{.04\textwidth}
\begin{minipage}{.48\textwidth}
\includegraphics[width=1.0\textwidth]{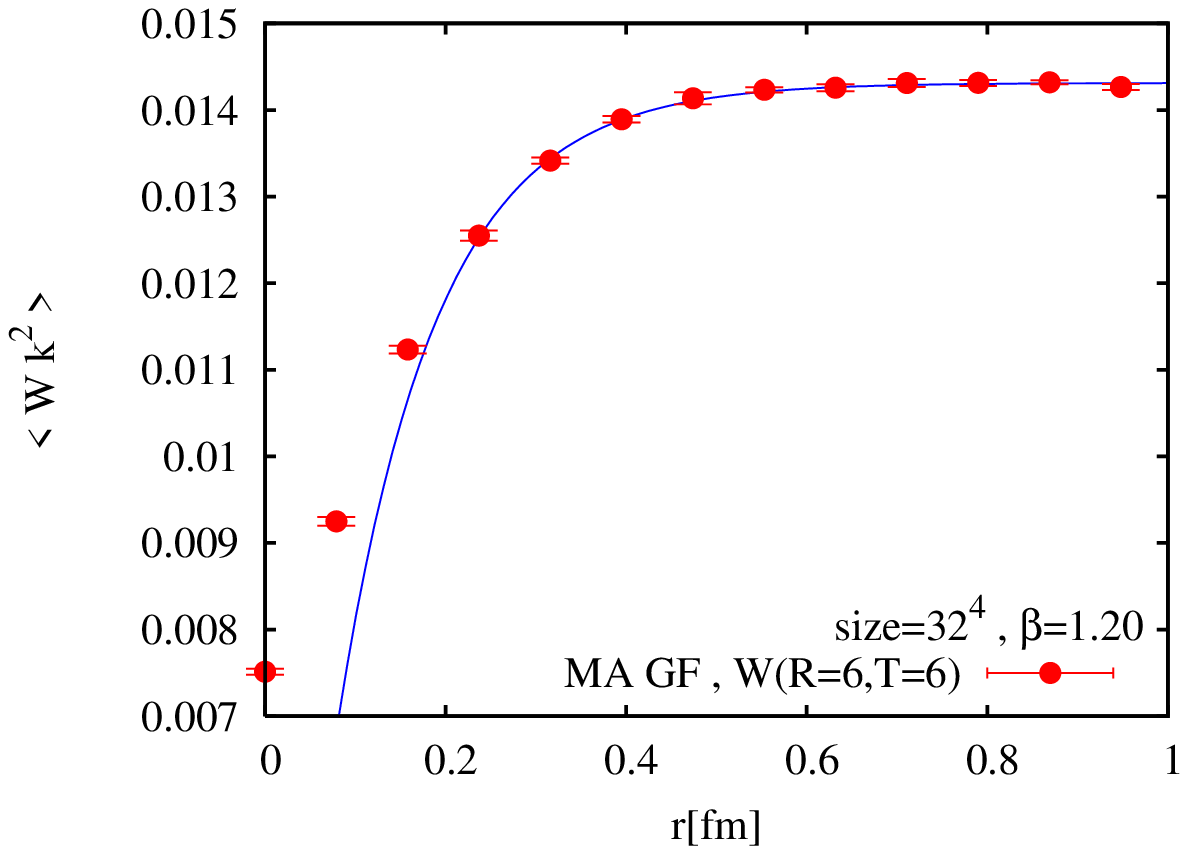}
\caption{
\label{w-mono}
 The correlation between the $R\times T= 6\times 6$ Wilson loop and the
square monopole density in the MA gauge. The solid line denotes the best
 exponential fit.}
\end{minipage}
\end{figure}

The numerical
measurements~\cite{ref:ambiguity} of the local correlation between
monopoles and the quantity $R_l$ revealed that the $A^+A^-$
condensate is enhanced on monopoles. Moreover, according to Fig.~\ref{k-AA}
the correlation between the $A^+A^-$ condensate and the monopole is short
ranged with the correlation length $\zeta_{\mathrm{cond}} \approx 0.06$~fm.
Note that $\zeta_{\mathrm{cond}} \approx \zeta_{\mathrm{Action}} \approx 0.05$fm where
the $\zeta_{\mathrm{Action}}$ is the scale of correlations between
the monopole density and the SU(2) action~\cite{ref:anatomy}.

\subsubsection{Correlations of monopoles and condensates with chromoelectric strings}

Let us then derive the coherence length in the MA gauge. The
correlations between the Wilson loop and the monopole density is
plotted in Fig.~\ref{w-mono}. The static quarks are represented by
the Wilson loop $W(R,T)$. The measurements of correlations are
mainly done on the perpendicular plane at the midpoint between the
quark pair. An exponential fit of the correlation function
provides the correlation length of the vacuum~$\xi$ according to
our discussion above.

\begin{figure}[htb]
\begin{minipage}{.48\textwidth}
\includegraphics[width=1.0\textwidth]{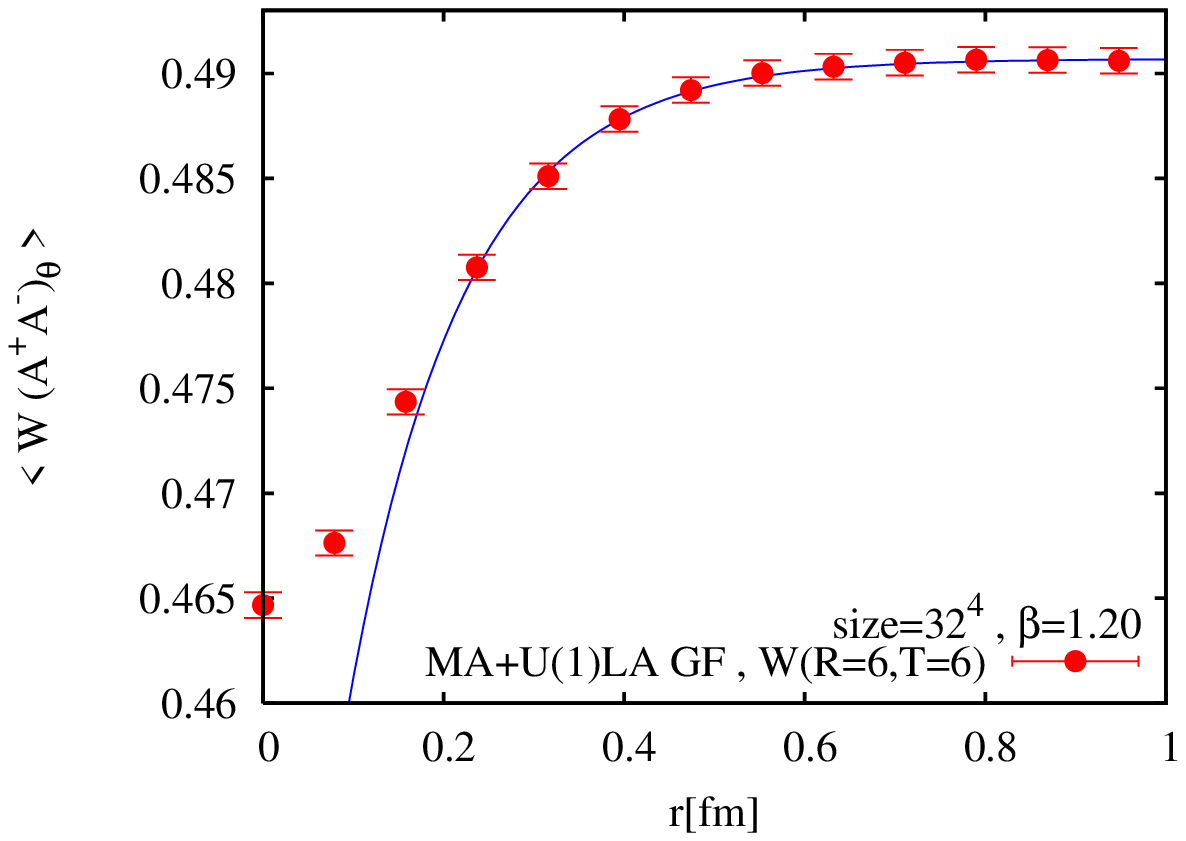}
\caption{\label{w-AA_theta} The correlation between the $R\times T= 6\times 6$ Wilson loop and
 the $A^+A^-_\theta$ in the MA $+$ U1LA gauge. The solid line denotes
 the best exponential fit.}
\end{minipage}
\hspace{.04\textwidth}
\begin{minipage}{.48\textwidth}
\includegraphics[width=1.0\textwidth]{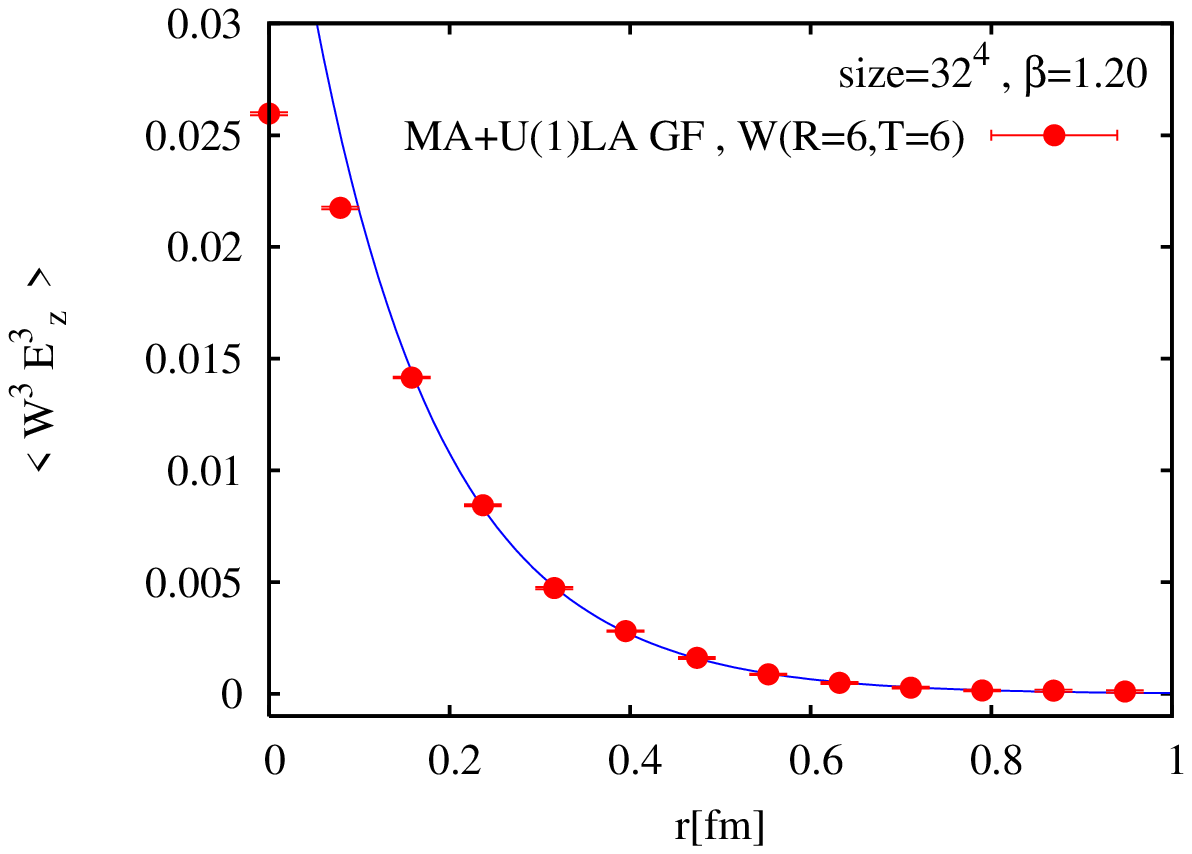}
\caption{\label{fig_1_ma} The non-Abelian $\vec{E}$
 electric field profile in the MA $+$ U1LA gauge obtained with the use of the $R\times T= 6\times 6$
 Wilson loop. The solid line denotes the best exponential fit.}
\end{minipage}
\end{figure}

Since the monopoles and the dimension-2 condensate are strongly
correlated, the coherence length can also be calculated from the
correlations between the Wilson loop and the dimension-2 quantity
${[A^+A^-(s)]}_\theta=\sum_\mu
\{[\theta_\mu^1(s)]^2+[\theta_\mu^2(s)]^2\}$ which uses the angles
$\theta_\mu(s)$ given by the relation $U_\mu(s)=
\exp(i\theta_\mu^a(s)\sigma^a)$. The quantity
${[A^+A^-(s)]}_\theta$ (measured in the MA+U1LA gauge) is
identical in naive continuum limit to the quantity $A^+A^-(s)$
defined in the MA gauge. The corresponding correlation function is
shown in Fig.~\ref{w-AA_theta}. We find the coherence lengths
determined by the use of the quantities ${[A^+A^-]}_\theta$ and
$k^2$  coincide
 within the error bars.


To derive the penetration length we study the correlation of the
non-Abelian electric fields which are defined from $1\times 1$ plaquette
$U_{\mu\nu}(s)=U^0_{\mu\nu}(s)+iU^a_{\mu\nu}(s)\sigma^a$. A
typical example is shown in Fig.~\ref{fig_1_ma}. Note that
electric fields perpendicular to the $Q\bar{Q}$ axis are found to
be negligible. An exponential fit of this correlation function
provides the penetration length of the vacuum, $\lambda$.

\begin{figure}[htb]
\begin{minipage}{.48\textwidth}
\includegraphics[width=1.0\textwidth]{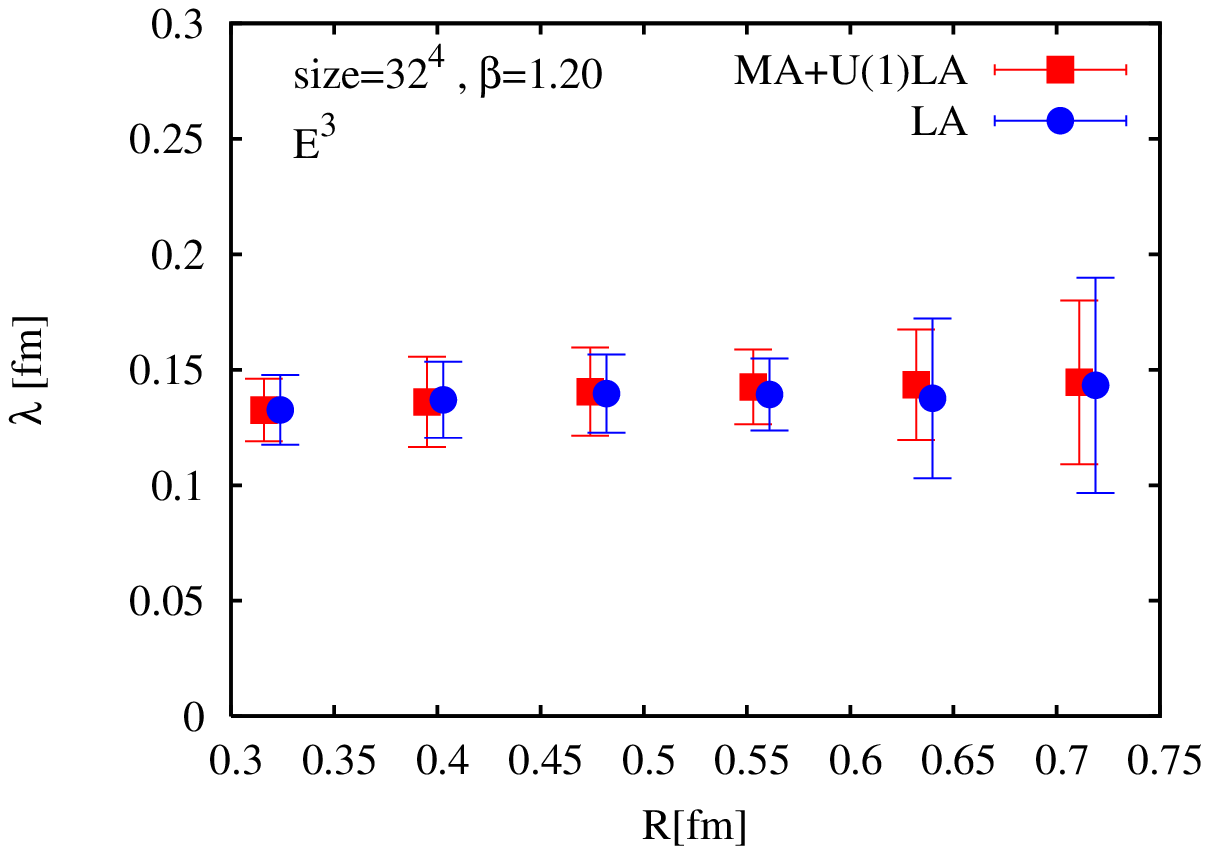}
\caption{\label{comp_penetration}
The penetration lengths of the non-Abelian electric field in the Landau gauge and
in the MA $+$ U1LA gauge for various $R$.}
\end{minipage}
\hspace{.04\textwidth}
\begin{minipage}{.48\textwidth}
\includegraphics[width=1.0\textwidth]{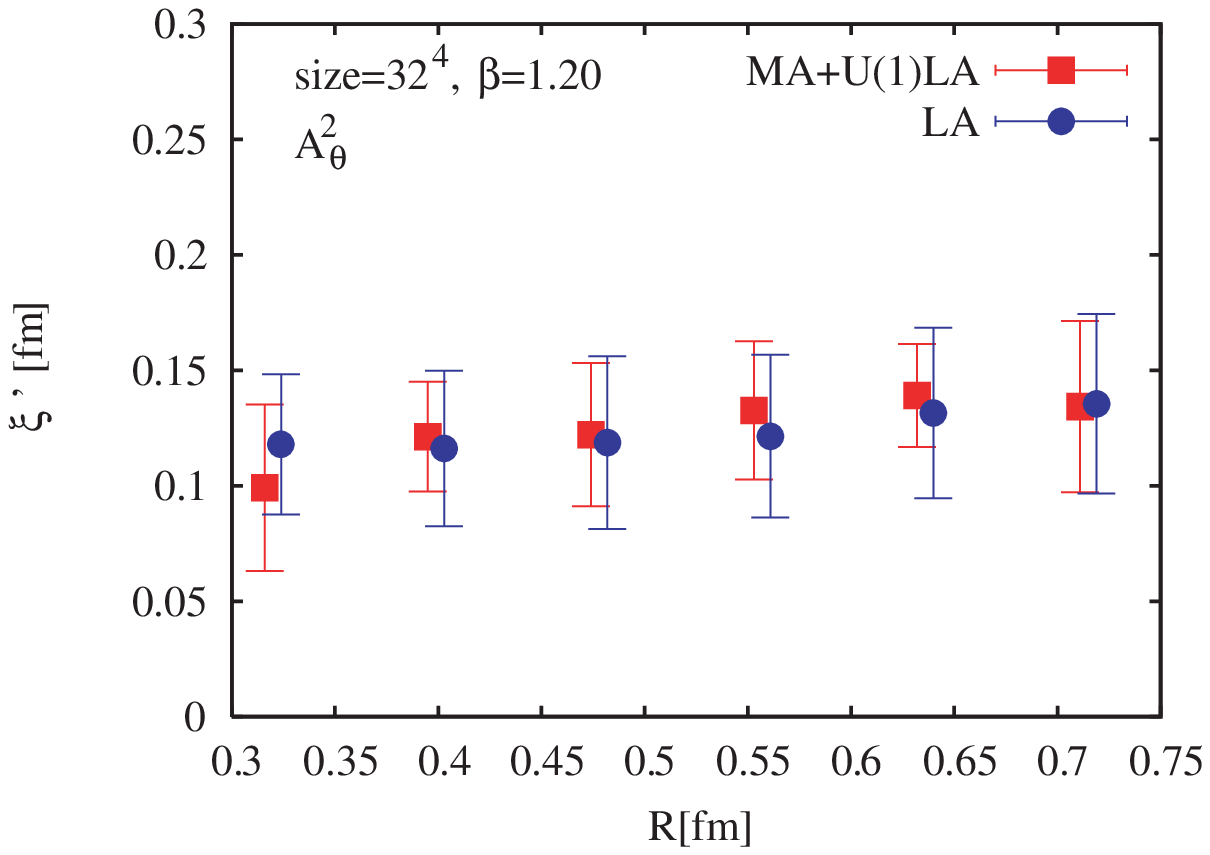}
\caption{\label{comp_coherence}
The coherence lengths of the dimension 2 gluon operator in the Landau gauge and in
the MA $+$ U1LA gauge for various $R$.}
\end{minipage}
\end{figure}

\subsection{LA gauge}

In the LA gauge the functional $\sum_{s,\mu} Tr [ U_{\mu}(s) +
U_{\mu}^{\dagger}(s)]$ is maximized with respect to all gauge
transformations. Similarly to the case of the MA gauge, we find
the coherence length from the measurement of the correlation between the chromoelectric string and
the dimension-2 operator
\begin{eqnarray}
A^2(s) \equiv \sum_{\mu=1}^{4} \sum_{a=1}^{3}\left(\theta_\mu^a(s)\right)^2\,.
\label{AS1}
\end{eqnarray}
Indeed, in the LA gauge the operator $A^2(s)$ (or its square-root)
is physically relevant and may have information about properties of a dual Higgs field
characterizing the QCD vacuum.

The $A^2(s)$ profile around the string in the LA gauge is very
similar to that shown in Fig.~\ref{w-AA_theta} for the MA gauge.
This is very exciting, since the behavior of the profile is just
what we expect from a profile of a Higgs field. The coherence
length is obtained from the exponential fitting of the correlation
function similarly to the MA case.

To derive the penetration length we study the correlation of the
Wilson loops with electric fields $E_{Ai}^a$ defined in the LA
gauge. The correlations in the LA gauge are very similar to the
case of the MA gauge shown in Fig.~\ref{fig_1_ma}.

\subsection{The vacuum type: comparison between MA gauge and LA gauge}

In order to study the gauge-(in)dependence of the dual
superconductor picture, we show in Fig.~\ref{comp_penetration} the
penetration lengths determined in the MA $+$ U1LA gauge and in the
LA gauge. We also plot the coherence lengths in
Fig.~\ref{comp_coherence}. {}From these figures, we observe that
the coherence and correlation lengths calculated in different
gauges coincide with each other. Note that we measure the
correlation between the gauge-invariant Wilson loop and the
gauge-invariant (``gauge-singlet'') pieces of the a gauge-variant
operators $A^2$. Since the $A^2$ operators are defined with
respect to different gauges their gauge-singlet parts are
non-local and different. Therefore the observed equivalence of the
correlations lengths is a non-trivial fact.

The Ginzburg-Landau parameter ({\it i.e.}, the ratio of penetration
length and the coherence length) determines the type of the SU(2) vacuum.
According to our measurements
\begin{eqnarray}
\kappa_{MA}&=&1.04 (\pm 0.07 \rm{statistic}) (\pm 0.1 \rm{systematic})\qquad [\mbox{MA gauge}]\,,\nonumber\\
\kappa_{LA}&=&1.04 (\pm 0.05 \rm{statistic}) (\pm 0.1 \rm{systematic})\qquad [\mbox{LA gauge}]\,. \nonumber
\end{eqnarray}

\section{Conclusions}

\begin{enumerate}
  \item The coherence lengths of the vacuum of the SU(2)
 gluodynamics in the MA gauge can {\it equivalently} be fixed either (i) from the correlations
 between the Wilson loops and the monopole density, or (ii) from the correlations
 between the Wilson loops and the dimension 2 operators.
  \item   The coherence lengths  measured in the MA gauge and in the LA gauge are the same.
  \item   The penetration lengths measured in the MA gauge and in the LA gauge are the same.
  \item  The type of the vacuum in both gauges is determined to be near
   the border between type 1 and type 2. The Ginzburg-Landau parameters in both gauges
   coincide with each other.
\end{enumerate}

\end{document}